\documentclass[12pt]{article}
\usepackage{amsmath,amsfonts,amssymb}  

\makeatletter
\def\numberbysection{\@addtoreset{equation}{section}
    \def\theequation{\thesection.\arabic{equation}}}
\makeatother

\numberbysection

\newcommand{\be}{\begin{eqnarray}}
\newcommand{\ee}{\end{eqnarray}}
\newcommand{\non}{\nonumber}

\def\bJ{{\mathbb J}}
\def\bR{{\mathbb R}}
\def\bL{{\mathbb L}}
\def\bQ{{\mathbb Q}}
\def\bC{{\mathbb C}}
\def\bH{{\mathbb H}}

\begin{document}

\begin{titlepage}
\strut\hfill ITP-Budapest Report 650
\vspace{.5in}
\begin{center}

\LARGE Yangian symmetry 
of boundary scattering in AdS/CFT\\
and the explicit form of bound state reflection matrices\\

\vspace{1in}
\large       L\'aszl\'o Palla  \footnote{palla@ludens.elte.hu
       } \\
       {\it Institute for Theoretical Physics, Roland E\"otv\"os University \\
          H-1117 P\'azm\'any s. 1/A, Budapest, Hungary}\\
\end{center}

\vspace{.5in}

\begin{abstract}
The reflection matrices of multi magnon bound states are  
obtained explicitely by 
exploiting the Yangian symmetry of boundary scattering on the
$Y=0$ maximal giant graviton brane.   
\end{abstract}

\end{titlepage}

\setcounter{footnote}{0}

\section{Introduction}

Since the fundamental two particle $S$ matrix of AdS/CFT \cite{St, Be1, AFZ} is determined 
by the symmetries (centrally extended $su(2|2)$\footnote{In this paper we
  focus on one copy of $su(2|2)$ symmetric $S$ and $R$ matrices, in the full
  AdS/CFT the $S$ and $R$ matrices are tensor products of two such copies.}) up to an overall scalar
factor the discovery that it also admits an interesting Yangian symmetry 
$Y(su(2|2))$ \cite{Be2} may seem to be relevant only from a mathematical point of
view. (Further investigations of Yangian symmetry in AdS/CFT can be found
e.g. in \cite{DNW}-\cite{Zw}). The power of the Yangian symmetry becomes manifest when one
tries to construct the $S$ matrices for the infinite tower of magnon bound
states \cite{CDO, AF}: ordinary 
symmetry considerations alone are not sufficient to fix the
form of the bound state $S$ matrix elements in general \cite{AF} 
and only the Yangian
symmetry is powerful enough to do this in case of the $Q$ bound state - 
$Q^\prime$ bound state scatterings ($Q,Q^\prime \ge 2$) \cite{AdLT}. 

The first steps to extend the Yangian symmetry to scattering in {\sl boundary} 
AdS/CFT are made in \cite{ANY, MKR}. In \cite{ANY} it is shown that the
remaining symmetry algebra of boundary scattering on the $Y=0$ brane is not
restrictive enough to make the reflection matrix of the $Q=2$ bound state
diagonal and leaves some elements of this matrix undetermined in contrast to
the reflection of the fundamental magnon, where the reflection matrix is
diagonal and is determined up to an overall factor. To remedy this situation
the authors of \cite{ANY} construct a conserved charge corresponding to a
generator of the boundary remnant of the bulk Yangian symmetry and show that
this extra conservation gives the missing equations leading to a solution of
the $Q=2$ bound state reflection matrix. 

In \cite{MKR} the structure of the Yangian symmetry of the $Y=0$ brane is
described in details, building on a series of earlier papers
\cite{DMS}-\cite{MK}. Using
the superspace formalism introduced in \cite{AF} they also construct the
reflection matrices of the fundamental ($Q=1$) and $Q=2$ bound state
reflections in terms of appropriate differential operators acting on 
superspace and combining this with the Yangian symmetry obtain the 
explicit form of
these $R$ matrices. 

The aim of this paper is to consider the reflection of a general $Q$ magnon
bound state ($Q>2$) on the $Y=0$ brane and to show that the boundary remnant
of the Yangian symmetry is powerful enough to to yield an explicit solution
even in this case. 

The paper is organized as follows: in the second section we review the 
superspace description of the $Q$ magnon bound states and discuss the general
structure of their reflection matrices with the outcome that $Q-1$ functions
in those matrices remain undetermined by ordinary 
symmetry considerations. In section 3 
we describe briefly the Yangian symmetry of the $Y=0$ brane and derive the
explicit solution for reflection matrices of the $Q$ magnon bound states. In 
section 4 we summarize our results and discuss also briefly the question of
Yangian symmetry for the bound states of the {\sl mirror} model \cite{AFtba}.       

\section{Magnon bound states and the structure of their reflection matrices}\label{sec:Magnonbound}

In this section we collect the necessary ingredients to describe the bound
states of $Q$ fundametal magnons as well as derive the general structure of
the reflection matrices describing the reflections of these bound states on 
the $Y=0$ brane.  

\subsection{$Q$ magnon bound state representation}\label{subsec:Qmagnon}

The centrally extended $su(2|2)$ algebra consists of the rotation
generators $\bL_{a}^{\ b}$, $\bR_{\alpha}^{\ \beta}$, the supersymmetry
generators $\bQ_{\alpha}^{\ a}$, $\bQ_{a}^{\dagger \alpha}$, and the
central elements $\bC\,, \bC^{\dagger}\,, \bH$.  Latin indices $a\,, b\,, 
\ldots$ take values $\{1\,, 2\}$, while Greek indices $\alpha\,, 
\beta\,, \ldots$ take values $\{3\,, 4\}$. These generators have the
following nontrivial commutation relations \cite{Be1, AFZ}
\be
\left[ \bL_{a}^{\ b}\,, \bJ_{c} \right] &=& \delta_{c}^{b} \bJ_{a} - 
\frac{1}{2} \delta_{a}^{b} \bJ_{c}\,, \quad 
\left[ \bR_{\alpha}^{\ \beta}\,, \bJ_{\gamma} \right] =
\delta_{\gamma}^{\beta} \bJ_{\alpha} - 
\frac{1}{2} \delta_{\alpha}^{\beta} \bJ_{\gamma}\,, \non  \\
\left[ \bL_{a}^{\ b}\,, \bJ^{c} \right] &=& -\delta_{a}^{c} \bJ^{b} + 
\frac{1}{2} \delta_{a}^{b} \bJ^{c}\,, \quad 
\left[ \bR_{\alpha}^{\ \beta}\,, \bJ^{\gamma} \right] =
-\delta_{\alpha}^{\gamma} \bJ^{\beta} +
\frac{1}{2} \delta_{\alpha}^{\beta} \bJ^{\gamma}\,, \non \\
\Big\{\bQ_{\alpha}^{\ a}\,, \bQ_{\beta}^{\ b}\Big\}&=& 
\epsilon_{\alpha \beta}\epsilon^{a b} \bC \,, \quad 
\Big\{\bQ_{a}^{\dagger \alpha}\,, \bQ_{b}^{\dagger \beta} \Big\}=
\epsilon^{\alpha \beta}\epsilon_{a b} \bC^{\dagger} \,, \non \\
\Big\{\bQ_{\alpha}^{\ a}\,, \bQ_{b}^{\dagger \beta} \Big\} &=& \delta_{b}^{a} 
\bR_{\alpha}^{\ \beta}+ \delta_{\alpha}^{\beta} \bL_{b}^{\ a} 
+ \frac{1}{2} \delta_{b}^{a} \delta_{\alpha}^{\beta} \bH \,,
\label{symmetryalgebra}
\ee
where $\bJ_{i}$ ($\bJ^{i}$) denotes any lower (upper) index of a generator,
respectively.

The Q magnon bound states form a $4Q$-dimensional atypical totally
symmetric representation of $su(2|2)$ \cite{CDO, AF}.  Using the convenient
superspace formalism of \cite{AF}, the $su(2|2)$ generators can be
represented by differential operators acting on a vector space of polynomials 
 built from two bosonic ($w_{a}$) and two fermionic 
($\theta_{\alpha}$) variables, as follows:
\be
\bL_{a}^{\ b} &=& 
w_{a}\frac{\partial}{\partial w_{b}}
-\frac{1}{2}\delta^{b}_{a}w_{c}\frac{\partial}{\partial w_{c}} \,, 
\qquad \qquad
\bR_{\alpha}^{\ \beta} = 
\theta_{\alpha}\frac{\partial}{\partial \theta_{\beta}}
-\frac{1}{2}\delta^{\beta}_{\alpha}\theta_{\gamma}
\frac{\partial}{\partial \theta_{\gamma}} \,, \non \\
\bQ_{\alpha}^{\ c} &=&
a\, \theta_{\alpha}\frac{\partial}{\partial w_{c}}
+ b\, \epsilon^{c b}\epsilon_{\alpha \beta} w_{b} 
\frac{\partial}{\partial \theta_{\beta}} \,, \qquad
\bQ_{a}^{\dagger \alpha} =
d\, w_{a} \frac{\partial}{\partial \theta_{\alpha}}
+ c\, \epsilon_{a b}\epsilon^{\alpha \beta}
\theta_{\beta}
\frac{\partial}{\partial w_{b}} \,, \non \\
\bC &=&
a b\, \left(w_{a}\frac{\partial}{\partial w_{a}}+
\theta_{\alpha}\frac{\partial}{\partial \theta_{\alpha}}\right)\,,
\qquad \quad
\bC^{\dagger} =
c d\, \left(w_{a}\frac{\partial}{\partial w_{a}}+
\theta_{\alpha}\frac{\partial}{\partial \theta_{\alpha}}\right)\,, \non \\
\bH &=&
(a d + b c) \left(w_{a}\frac{\partial}{\partial w_{a}}+
\theta_{\alpha}\frac{\partial}{\partial \theta_{\alpha}}\right)\,.
\label{superspacerep}
\ee 
The $a,b,c,d$ parameters (satisfying $ad-bc=1$) are functions of the $Q$ magnon 
bound state's momentum $p$ \cite{Be1, AF}:
\be
a = \sqrt{\frac{g}{2Q}}\eta\,, \quad 
b = \sqrt{\frac{g}{2Q}}\frac{i}{\eta}\left(\frac{x^{+}}{x^{-}}-1\right)\,, \quad 
c= -\sqrt{\frac{g}{2Q}}\frac{\eta}{x^{+}}\,, \quad 
d=\sqrt{\frac{g}{2Q}}\frac{x^{+}}{i \eta}\left(1 - \frac{x^{-}}{x^{+}}\right)\,,
\label{BulkParameters}
\ee
where (we follow the phase convention of \cite{AF}) 
\be
x^{+}+\frac{1}{x^{+}}-x^{-}-\frac{1}{x^{-}} = \frac{2Q i}{g}\,, \qquad 
\frac{x^{+}}{x^{-}} = e^{i p} \,,\qquad \eta = e^{i p/4}\sqrt{i(x^{-}-x^{+})}  \,.
\label{xpm}
\ee 
We decompose the $4Q$ dimensional representation space, ${\cal V}^Q(p)$, 
as $4Q=(Q+1)+(Q-1)+Q+Q$ 
and parameterize the sub-spaces as
\be
Q+1\rightarrow\vert j\rangle^1 &=&
\frac{w_1^{Q-j}w_2^j}{\sqrt{(Q-j)!j!}},\qquad j=0,\dots Q \non \\  
Q-1\rightarrow\vert j\rangle^2 &=&
\frac{w_1^{Q-2-j}w_2^j}{\sqrt{(Q-2-j)!j!}}\theta_3\theta_4,\qquad j=0,\dots Q-2
\non \\ 
Q\rightarrow\vert j\rangle^3 &=&
\frac{w_1^{Q-1-j}w_2^j}{\sqrt{(Q-1-j)!j!}}\theta_3,\qquad j=0,\dots Q-1 \non \\
Q\rightarrow\vert j\rangle^4 &=&
\frac{w_1^{Q-1-j}w_2^j}{\sqrt{(Q-1-j)!j!}}\theta_4,\qquad j=0,\dots Q-1 \,.
\label{decomp}
\ee
(Note, that for the fundamental magnon, $Q=1$, the second subspace is absent
and the $4$ dimensional representation is parameterized as $4=2+1+1$ with 
$\vert 0\rangle^1=w_1$, $\vert 1\rangle^1=w_2$, $\vert 0\rangle^3=\theta_3$, 
$\vert 0\rangle^4=\theta_4$). 

\subsection{The structure of the reflection matrix}\label{subsec:Rmatrix}  

Next we investigate to what extent  the remaining symmetries of the
$Y=0$ brane restrict the
reflection matrix of the Q magnon bound states. These remaining symmetries
form an $su(2|1)$ sub-algebra \cite{HM}, consisting of the  
generators
\be
\bL_{1}^{\ 1}\,, \quad  \bL_{2}^{\ 2}\,, \quad\bH \,, \quad 
\bR_{\alpha}^{\ \beta}\,, \quad  \bQ_{\alpha}^{\ 1}\,, \quad  
\bQ_{1}^{\dagger \alpha} \,.
\label{su12gens}
\ee    
To describe the reflection matrix we follow \cite{ANY} , \cite{MKR} and define 
a boundary vacuum state $\vert 0\rangle_B$ corresponding to a trivial vector 
space ${\cal V}(0)$ annihilated by all $su(2|1)$ generators. This makes it
possible to define the (super-space) $R$-matrix for the reflection of bulk  
magnon bound states as an operator acting on the tensor product spaces 
\be
R(p):\quad {\cal V}^Q(p)\otimes{\cal V}(0)\rightarrow {\cal
  V}^Q(-p)\otimes{\cal V}(0)
\label{Rdef}
\ee
where the reflection matrix is given as a differential operator 
\be 
R(p)=\sum_i r_i(p)\Lambda_i \label{Rlambda}
\ee
acting on the super-space. Here $\Lambda_i$ span a basis of invariant
differential operators built from $w_a$, and $\theta_\alpha$. 
Since all $\bJ^i$ generators of $su(2|1)$, eq.(\ref{su12gens}), annihilate
$\vert 0\rangle_B$, on the tensor product space they have the coproducts
\cite{MKR}:
\be
\triangle (\bJ^i)=\bJ^i\otimes 1\,. \non
\ee
As a consequence requiring the reflections to respect the $su(2|1)$ symmetry
amounts to imposing the vanishing of the commutator $[\bJ^i,R]\vert
j\rangle^I$ $I=a,\alpha$.   

Since $\bL_1^{\ 1}$ ($\equiv -\bL_2^{\ 2}$) acts diagonally on the various 
sub-spaces 
\be
\bL_1^{\ 1}\vert j\rangle^1=\frac{1}{2}(Q-2j)\vert j\rangle^1,\quad 
\bL_1^{\ 1}\vert j\rangle^2=\frac{1}{2}(Q-2-2j)\vert j\rangle^2,\quad
\bL_1^{\ 1}\vert j\rangle^\alpha=\frac{1}{2}(Q-1-2j)\vert j\rangle^\alpha, \non
\ee
and $\bR_\alpha^{\ \beta} $ acts non trivially only for $\vert j\rangle^\gamma
$, and in particular
\be
\bR_4^{\ 3}\vert j\rangle^3=\vert j\rangle^4, \qquad
\bR_3^{\ 4}\vert j\rangle^4=\vert j\rangle^3, \non
\ee
requiring these \lq\lq bosonic'' symmetry generators to commute with $R$
restricts the form of the reflection matrix as
\be
R=\sum\limits_{l=0}^Q A_l\Lambda_{(1)}^l +\sum\limits_{l=0}^{Q-2}
B_l\Lambda_{(2)}^l + \sum\limits_{l=0}^{Q-1} C_l\Lambda_{(3)}^l 
+\sum\limits_{l=0}^{Q-2}D_l\Lambda_{(4)}^l
+\sum\limits_{l=0}^{Q-2}E_l\Lambda_{(5)}^l \,.
\label{Rsajat}
\ee 
Here the various differential operators are given as
\be
\Lambda_{(1)}^l=\frac{w_1^{Q-l}w_2^l}{(Q-l)!l!}\frac{\partial^{Q}}{\partial
  w_1^{Q-l}\partial w_2^l},\quad 
\Lambda_{(2)}^l=\frac{w_1^{Q-2-l}w_2^l}{(Q-2-l)!l!}\theta_3\theta_4
\frac{\partial^{Q-2}}{\partial
  w_1^{Q-2-l}\partial
  w_2^l}\frac{\partial^2}{\partial\theta_4\partial\theta_3},
\non 
\ee
\be
\Lambda_{(3)}^l=\frac{w_1^{Q-1-l}w_2^l}{(Q-1-l)!l!}\frac{\partial^{Q-1}}{\partial
  w_1^{Q-1-l}\partial
  w_2^l}\theta_\alpha\frac{\partial}{\partial\theta_\alpha} ,\non
\ee
\be 
\Lambda_{(4)}^l=\frac{w_1^{Q-2-l}w_2^l}{(Q-2-l)!l!}\theta_3\theta_4
\frac{\partial^{Q}}{\partial
  w_1^{Q-1-l}\partial w_2^{l+1}} ,\quad 
\Lambda_{(5)}^l=\frac{w_1^{Q-1-l}w_2^{l+1}}{(Q-2-l)!l!}
\frac{\partial^{Q-2}}{\partial
  w_1^{Q-2-l}\partial
  w_2^l}\frac{\partial^2}{\partial\theta_4\partial\theta_3}\,,
\non 
\ee
and $A_l$, $B_l$, $C_l$, $D_l$ and $E_l$ are $5Q-2$ (unknown) functions of
$p$. (For $Q=1$ the second, fourth and fifth sums are missing leaving only
$A_0$ $A_1$ and $C_0$ to be determined). The first three sums in
(\ref{Rsajat}) 
describe
diagonal reflections while the fourth and fifth ones describe off-diagonal 
reflections between multi-magnon polarizations in 
the $Q+1$ and $Q-1$ subspaces: this possibility arises as a result of the 
coinciding eigenvalues of the bosonic symmetry generators in the two
subspaces. 

To obtain equations for the unknown functions next we consider the
restrictions following from requiring also the fermionic generators to commute 
with reflections. For this we list the action of the fermionic generators
on the various sub-spaces 
\be
\bQ_\alpha^{\ 1}\vert j\rangle^1 &=& a\sqrt{Q-j}\vert j\rangle^\alpha,\qquad
\bQ_\alpha^{\ 1}\vert j\rangle^2=-b\sqrt{j+1}\vert j+1\rangle^\alpha, \non \\
 \bQ_\alpha^{\ 1}\vert
 j\rangle^\beta &=& \epsilon^{\alpha\beta}(a\sqrt{Q-1-j}\vert
 j\rangle^2+b\sqrt{j+1}\vert j+1\rangle^1), \non 
\ee
\be
\bQ_{1}^{\dagger \alpha}\vert j\rangle^1 &=&
c\sqrt{j}\epsilon^{\alpha\beta}\vert j-1\rangle^\beta ,\qquad 
\bQ_{1}^{\dagger \alpha}\vert j\rangle^2 = 
d\sqrt{Q-1-j}\epsilon^{\alpha\beta}\vert j\rangle^\beta , \non \\
\bQ_{1}^{\dagger \alpha}\vert j\rangle^\beta &=&
\delta^{\alpha\beta}(d\sqrt{Q-j}\vert j\rangle^1-c\sqrt{j}\vert j-1\rangle^2)
\,. \non
\ee
To make the subsequent equations simpler we introduce the notation
$\dot{f}\equiv f(-p)$ for any function $f(p)$. Using this and recalling
(\ref{Rdef}) 
the action of the
reflection matrix can be written as
\be
R\vert j\rangle^1 &=& A_j\dot{(\vert j\rangle^1)}+\sqrt{(Q-j)j}D_{j-1}\dot{(\vert
  j-1\rangle^2)} , \qquad j=0,\dots ,Q \non \\
R\vert j\rangle^2 &=& B_j\dot{(\vert j\rangle^2)}+\sqrt{(Q-1-j)(j+1)}E_j\dot{(\vert
  j+1\rangle^1)} , \qquad j=0,\dots ,Q-2 \non \\
R\vert j\rangle^\alpha &=& C_j\dot{(\vert j\rangle^\alpha)} .\qquad j=0,\dots ,Q-1 \non
\ee
The vanishing of the commutator on the four subspaces 
$[\bQ_\alpha^{\ 1},R]\vert j\rangle^I$
$I=a,\alpha$ leads to the following equations:
\be
aC_j &=& \dot{a}A_j-\dot{b}jD_{j-1},\qquad\qquad\qquad\ \   j=0,\dots ,Q-1,\non \\
bC_{j+1} &=& \dot{b}B_j-\dot{a}(Q-1-j)E_j,\qquad\quad  j=0,\dots ,Q-2, \non \\
\dot{a}C_j &=& aB_j+b(j+1)D_j, \qquad\quad\qquad j=0,\dots ,Q-2, \non \\
\dot{b}C_j &=& bA_{j+1}+a(Q-1-j)E_j, \quad\quad j=0,\dots ,Q-1, 
\label{Qequ} 
\ee 
while using $\bQ_1^{\dagger \alpha}$ instead of $\bQ_\alpha^{\ 1}$ gives
\be
cC_{j-1} &=& \dot{c}A_j+\dot{d}(Q-j)D_{j-1},\qquad\quad j=1,\dots ,Q,\non \\
dC_{j} &=& \dot{d}B_j+\dot{c}(1+j)E_j, \qquad\quad\quad\  j=0,\dots ,Q-2, \non \\
\dot{d}C_j &=& dA_j-cjE_{j-1}, \qquad\qquad\qquad j=0,\dots ,Q-1, \non \\
\dot{c}C_j &=& cB_{j-1}-d(Q-j)D_{j-1}, \quad\quad j=1,\dots ,Q-1. 
\label{Qbequ} 
\ee  
In both sets of equations there are altogether $4Q-2$ equations. However it is 
straightforward to show that in both sets there are $2(Q-1)$ relations between 
these equations, leaving in both sets $2Q$ independent equations. Since 
$\dot{a}/a=d/\dot{d}=e^{-ip/2}$ and
$\dot{c}/c=b/\dot{b}=-e^{ip/2}$ the two (independent) 
\lq\lq corner''
equations connecting $A_0$ to $C_0$ and $A_Q$ to $C_{Q-1}$ are identical in
(\ref{Qequ}) and in (\ref{Qbequ}); thus in the two sets there are only $4Q-2$
independent equations. This means that
$Q$ of the  $5Q-2$ unknown  functions is not determined by requiring the
symmetry transformations and reflections to commute. On physical grounds we
expect that one overall scalar factor in the reflection matrix is determined
by consideration going beyond the symmetries (unitarity, crossing, fusion
etc.); setting this overall scale to one still leaves $Q-1$ functions
undetermined. In this respect the reflection of $Q$ magnon bound states is
different from the reflection of fundamental magnons as was discovered in the
$Q=2$ case in \cite{ANY}. In the next section we show that invoking the
Yangian extension of the $su(2|1)$ symmetry provides the necessary extra 
equations even in the general case.   
   
\section{Yangian symmetry and the explicit form of the
  reflection matrix}\label{sec:YandR}

In this section -- following \cite{ANY} \cite{MKR} -- 
we describe in a nutshell the Yangian symmetry of the $Y=0$
brane and present the explicit solution for the reflection matrix of the $Q$
magnon bound state. 

\subsection{The Yangian of the $Y=0$ brane}

The Yangian extension ($Y({{\bf g}})$) 
of a {\sl bulk} Lie symmetry ${\bf g}$ is a deformation
of the universal enveloping algebra of the polynomial algebra ${\bf
  g}(u)$. It is generated by grade-1 \lq\lq Yangian`` generators $\hat{\bJ}^A$ 
besides the grade-0 generators $\bJ^A$ of ${\bf g}$. Their commutation
relations have the form 
\be
[\bJ^A,\bJ^B]=f^{AB}_{\ \ \ C}\bJ^C,\qquad
[\bJ^A,\hat{\bJ}^B]=f^{AB}_{\ \ \ C}\hat{\bJ}^C,
\non 
\ee
and must obey the Jacobi and Serre relations. To satisfy these in case of
${\bf g}=su(2|2)$ is not entirely straightforward as the Killing form of 
$su(2|2)$ is degenerate, but this problem may be circumvented and the explicit
form of $Y(su(2|2))$ - together with the coproducts of the Yangian generators 
- is known \cite{Be2}. (The coproducts are necessary to explore the action of
$Y$ on two particle states).

In constructing finite dimensional representations of $Y({\bf g})$ a crucial 
role is played by the \lq\lq evaluation representation'', where the grade-1
generators have the form
\be
\hat{\bJ}^A\vert u\rangle =-i\frac{g}{2}u\bJ^A\vert u\rangle \,.\non
\ee
It is shown in \cite{Be2} \cite{AdLT}  
that the (multi) magnon (bound) states are of this
form, where 
\be
u\equiv u(p)=\frac{1}{2}(x^++\frac{1}{x^+}+x^-+\frac{1}{x^-}) \,, 
\label{udef}
\ee
and $x^\pm$ are defined in (\ref{xpm}). In \cite{Be2} it is shown that the
fundamental magnon's $S$-matrix is constrained up to an overall phase by
imposing the invariance of the $S$-matrix under (the coproducts of) 
$Y(su(2|2))$. In the case of scattering of a $Q$ bound state and a $Q^\prime$
bound state ($Q,Q^\prime\geq 2$) the $su(2|2)$ 
symmetry algebra is not enough to fix all
elements of $S$; the necessary additional constraints may be obtained from the 
Yangian symmetry \cite{AdLT}.
   
As discussed in a series of papers \cite{DMS}, \cite{MKS}, \cite{MK} 
only a remnant (denoted as $Y({\bf
  h},{\bf g})$) 
of the bulk Yangian 
symmetry survives when the bulk theory is restricted by a boundary, which
although preserving integrability preserves 
 only a 
sub-algebra ${\bf h}\subset {\bf g}$ of the symmetry of the bulk fields. As turns
out $({\bf h},{\bf g})$ must form a symmetric pair
\be
{\bf g}={\bf h}+{\bf m},\quad [{\bf h},{\bf h}]\subset {\bf h},\quad  
[{\bf h},{\bf m}]\subset {\bf m},\quad [{\bf m},{\bf m}]\subset {\bf h},\non
\ee
and $Y({\bf h},{\bf g})$ is generated by $(\bJ^i,\tilde{\bJ}^p)$, where $i(j,k)$
run over the ${\bf h}$ indeces and $p(q,r)$ over the ${\bf m}$ indeces, and
\be
\tilde{\bJ}^p =\hat{\bJ}^p +\frac{1}{2}f^p_{qi}\bJ^q\bJ^i\,.
\non
\ee
(The extra twisting represented by the second term is necessary to guarantee
that products of bulk and boundary states still represent $Y({\bf
  h},{\bf g})$). 

In case of the $Y=0$ brane the generators of ${\bf h}$ are given in 
(\ref{su12gens}) and the subspace ${\bf m}$ is generated by
\be
\bL_2^{\ 1},\quad \bL_1^{\ 2},\quad \bQ_\gamma^{\ 2},\quad
\bQ_2^{\dagger\ \gamma},\quad \bC,\quad \bC^\dagger \,. \non 
\ee
(It is straightforward to check that they indeed form a symmetric pair). Now 
one can readily construct the Yangian generators $\tilde{\bJ}^p$ and their 
coproducts exploiting that all symmetry generators annihilate the boundary
vacuum $\vert 0\rangle_B$; their explicit form is given in \cite{MKR}. In this
paper we use only one of them, 
\be
\tilde \bQ \otimes 1\equiv \triangle\tilde{\bL}_2^{\ 1} = \left( \hat \bL_{2}^{\ 1} + \frac{1}{2}\left( 
\bL_{2}^{\ 1} \bL_{1}^{\ 1} - \bL_{2}^{\ 1} \bL_{2}^{\ 2}
- \bQ_{2}^{\dagger \gamma} \bQ_{\gamma}^{\ 1} \right)\right)\otimes 1 \,,
\label{tildeQ}
\ee
the same one introduced in \cite{ANY}.      

\subsection{The explicit form of the reflection matrix}

We obtain equations supplementing (\ref{Qequ},\ref{Qbequ}) by imposing the vanishing 
of the commutator between $\tilde{\bQ}$ and $R$. To implement these we need
the action of $\tilde{\bQ}$ on the various subspaces:
\be
\tilde{\bQ}\vert j\rangle^1 &=& \sqrt{(Q-j)(j+1)}\left(
-i\frac{g}{2}u+\frac{Q}{2} -j-ad\right) \vert
j+1\rangle^1-\sqrt{(Q-j)(Q-1-j)}ac\vert j\rangle^2 ,\non \\
\tilde{\bQ}\vert j\rangle^2 &=& \sqrt{(Q-2-j)(j+1)}\left(
-i\frac{g}{2}u+\frac{Q-2}{2} -j+bc\right) \vert
j+1\rangle^2 \non \\ &+& \sqrt{(j+1)(j+2)}bd\vert j+2\rangle^1 ,\non \\
\tilde{\bQ}\vert j\rangle^\beta &=& \sqrt{(Q-1-j)(j+1)}\left(
-i\frac{g}{2}u+\frac{Q-1}{2} -j-\frac{1}{2}\right) \vert
j+1\rangle^\beta \,.\label{Qtildequ}
\ee
Since both $R$ and $\tilde{\bQ}$ act diagonally on the fermionic subspaces
$\vert j\rangle^\alpha$ requiring $[\tilde{\bQ},R]\vert j\rangle^\gamma=0$
gives $Q-1$ equations 
\be
C_{j+1}=\Phi(j)C_j,\quad j=0,\dots ,Q-2,\qquad {\rm where}\quad
\Phi(j)=\frac{i\frac{g}{2}u+\frac{Q}{2}-j-1}{-i\frac{g}{2}u+\frac{Q}{2}-j-1}.
\label{Cegyenlet}
\ee
These equations determine all the $C_j$ in terms of $C_0$:
\be
C_{j+1}=C_0\prod\limits_{l=0}^j\Phi(l),\qquad j=0,\dots ,Q-2 \,.
\label{Cmegold}
\ee
The structure of the set of solutions $\{ C_0,\ C_1,\ \dots C_{Q-1}\}$ 
depends on whether $Q$ is
even or odd. For $Q$ even, $l_0=\frac{Q-2}{2}$ is integer, and exploiting 
\be
\Phi(Q-2-l)=\frac{1}{\Phi(l)},\qquad {\rm and}\quad \Phi(l_0)=-1, \non
\ee
one can show that that the set of $C$-s has the form
\be
\{ C_0,\quad C_1,\quad\dots ,\quad C_{l_0},\quad -C_{l_0},\quad\dots ,\quad-C_1,\quad -C_0\}\,;\label{Qps}
\ee  
while for $Q$ odd, $Q=2r+1$, exploiting 
\be
\Phi(r-k)=\frac{1}{\Phi(r+k-1)},\qquad k=1,\dots ,r \non 
\ee
one obtains that the set of $C$-s have the form
\be
\{ C_0,\quad C_1,\quad\dots ,\quad C_{r-1},\quad C_{r},\quad C_{r-1},\quad
\dots ,\quad C_1,\quad C_0\}\,. \label{Qptl}
\ee 
For $Q=2$, $l_0=0$, and (\ref{Qps}) gives $C_1=-C_0$, which is consistent with 
\cite{ANY}. 

Now one can use these explicitely known $C_j$-s in eq.(\ref{Qequ},
\ref{Qbequ}) to determine the remaining unknown functions. We choose the
normalization $A_0=1$ (since $\vert 0\rangle^1$ has the highest $\bL_1^{\ 1}$
value), using this in the \lq\lq corner equations'' leads to 
\be
C_0=A_0\frac{d}{\dot{d}}=e^{-ip/2},\qquad
A_{Q}=\frac{c}{\dot{c}}C_{Q-1}=e^{-ip}\prod\limits_{l=0}^{Q-2}\Phi(l) =(-1)^Q
e^{-ip}.
\label{sol1}
\ee
The other unknown coefficients in the reflection matrix are found to be given
by 
\be
A_{j+1} &=& \left( \prod\limits_{l=0}^{j-1}\Phi(l)\right)
\frac{(Q-1-j)\Phi(j)x^+-(j+1)/x^+}{(Q-1-j)x^++(j+1)/x^-}, \non \\ 
B_{j} &=& \left( \prod\limits_{l=0}^{j-1}\Phi(l)\right) 
\frac{(Q-1-j)x^--(j+1)\Phi(j)/x^-}{(Q-1-j)x^++(j+1)/x^-}, \label{thesol} \\
E_{j} &=& -D_j=e^{-ip/2}\left( \prod\limits_{l=0}^{j-1}\Phi(l)\right) 
\frac{x^+\Phi(j)+x^-}{x^+x^-(Q-1-j)+(j+1)},\quad j=0,\dots Q-2\,,\non 
\ee
where, for $j=0$, $\prod\limits_{l=0}^{j-1}\Phi(l)=1$ is understood. The
expressions appearing in eq.(\ref{Cmegold}, \ref{sol1}) and (\ref{thesol})
constitute the explicit form of functions defining 
the reflection matrix of the $Q$ magnon bound
state (apart from the overall scalar factor) 
and they represent the main result of this paper. 

For $Q=2$ this solution has the form: $A_0=1$ and 
\be
C_0=e^{-ip/2}=-C_1,\qquad A_2=e^{-ip},\qquad  
A_1=-\frac{x^++\frac{1}{x^+}}{x^++\frac{1}{x^-}},\non \\ 
B_0=\frac{x^-+\frac{1}{x^-}}{x^++\frac{1}{x^-}},\qquad\qquad
E_0=-D_0=e^{-ip/2}\frac{x^--x^+}{1+x^+x^-}, \non   
\ee
which, recalling eq.(\ref{Rsajat}), is nothing but $R_{AN}(-p)\equiv
R_{AN}^{-1}(p)$, with $R_{AN}$
being the reflection matrix found in \cite{ANY}. (This difference follows from 
the difference between
our definition of the reflection matrix (\ref{Rdef}), and eq.(3.10) in
\cite{ANY}). 

It is straightforward to check that the solution given by 
eq.(\ref{Cmegold}, \ref{sol1}) and (\ref{thesol}) satisfies the unitarity 
constraint
\be
R(-p)R(p)=1.\label{Runitary}
\ee
Indeed the \lq\lq diagonal'' part of the reflection matrix -i.e. the one
determined by $A_0$, $A_Q$ and $C_j$ satisfy this as a consequence of
$\dot{\Phi(l)}=1/\Phi(l)$, while on $\vert j\rangle^1$, $\vert j\rangle^2$ 
(\ref{Runitary}) 
  gives four equations between $A_{j+1}$, $D_j$, $E_j$, $B_j$ and their \lq\lq
  dotted'' versions, which are found to be satisfied.  

We used only the fermionic sub-space component of the commutator
$[\tilde{\bQ},R]\vert j\rangle^\gamma$ 
to determine the functions characterizing the reflection matrix. For
consistency $[\tilde{\bQ},R]\vert j\rangle^1$ and $[\tilde{\bQ},R]\vert
j\rangle^2$ should also vanish. It is straightforward to write explicitely the
equations these requirements impose on $A_{j+1}$, $D_j$, $E_j$, $B_j$ and we
checked (albeit sometimes only numerically) that the solution given by 
eq.(\ref{Cmegold}, \ref{sol1}) and (\ref{thesol}) satisfies all of them.  

The eventual verification of our solution is to show that it solves the
boundary Yang-Baxter equation. Denoting by $R_Q$ ($R_{Q^\prime}$) the
reflection matrix of the $Q$ ($Q^\prime$) magnon bound states and by
$S_{QQ^\prime}$ their bulk $S$-matrix, the boundary Yang-Baxter equation has
the schematic form:
\be
R_Q(p)S_{Q^\prime Q}(q,-p)R_{Q^\prime}(q)S_{QQ^\prime}(-p,-q)=
S_{Q^\prime Q}(q,p)R_{Q^\prime}(q)S_{QQ^\prime}(p,-q)R_Q(p)\,.\label{bYB}
\ee
Although the explicit form of $S_{QQ^\prime}$ is known \cite{AdLT}, because of the
complexity of these expressions, the verification of (\ref{bYB}) is beyond the
scope of the present paper.  

For physical applications one needs the explicit form of the overall scalar 
factor we ignored so far. In fact this scalar factor is known: for the
fundamental magnon, $Q=1$, an equation for this factor is derived from
unitarity and crossing considerations in \cite{HM}, this equation is solved 
in \cite{CC}; finally using this solution as input the scalar factor of the
$Q$ magnon bound state is determined by the fusion method in \cite{ABR}
exploiting that the bound state's component with the highest value of
$\bL_1^{\ 1}$ scatters and reflects diagonally. 

\section{Summary and discussion}

In this paper the reflection of multi magnon bound states on the
$Y=0$ maximal giant graviton brane is investigated. It is shown that the
reflection matrices of the $Q$ magnon bound states can be described as 
appropriate linear combinations of projectors and 
are characterized in terms
of $5Q-2$ unknown functions. $Q-1$ of these functions  
remain undetermined by ordinary
symmetry considerations, i.e. by requiring 
the surviving $su(2|1)$ generators to commute with reflections. Invoking the 
Yangian extension of the $su(2|1)$ symmetry of the $Y=0$ brane solves the
problem: requiring the same Yangian generator that is used in \cite{ANY} to
commute with reflections gives precisely $Q-1$ additional equations such that 
the whole system admits a consistent solution. These explicit reflection
matrices - when augmented by the overall scalar factor available in the
literature \cite{ABR} - can be used as starting points to analyze the boundary
finite size effects for magnon bound states.      
   
In \cite{MKR} also a \lq\lq toy model'' boundary is discussed 
together with its
Yangian symmetry. In this model the hypothetical boundary
breaks $su(2|2)$ down to $\tilde{\bf h}=su(1|2)$ generated by 
\be
\bR_3^{\ 3}=\frac{1}{2}(\theta_3\frac{\partial}{\partial\theta_3}-
\theta_4\frac{\partial}{\partial\theta_4}),\quad \bL_a^{\ b},\quad 
\bQ_3^a,\quad \bQ_a^{\dagger 3},\quad\bH .\label{su12}
\ee
In \cite{MKR} it is shown that the corresponding Yangian extension 
$Y(\tilde{\bf h},su(2|2))$ is in a certain sense redundant or trivial, since
imposing only the $\tilde{\bf h}$ symmetry on the reflection matrix 
$R$, defined
according to (\ref{Rdef}), (\ref{Rlambda}) makes $R$ diagonal and determines
its elements (up to an overall factor)
in contrast to the physical case discussed in this paper. Note that the
essential difference between ${\bf h}$, generated by (\ref{su12gens}) and 
$\tilde{\bf h}$, generated by (\ref{su12}) is that in the first case the
unbroken bosonic generators are built from the fermionic while in the second
from the bosonic parameters. 

Our remark is that $\tilde{\bf h}$ and this second Yangian 
play a natural role for the bound states of the {\sl mirror model}
\cite{AFtba} obtained by a double Wick rotation from the physical one. 
The bound states of $Q$
fundamental {\sl mirror} magnons form the $4Q$ dimensional completely
{\sl antisymmetric} representation of $su(2|2)$ \cite{AFtba}
that can be described in the
super-space formalism in terms of the fermionic $\theta_1,\theta_2$ and bosonic 
$w_3,w_4$ parameters, i.e. in the mirror model one has to make the
$w\leftrightarrow \theta $ substitutions. Now making these changes in the
generators spanning $su(2|1)$ in (\ref{su12gens}) leads - after the trivial 
index change $(1,2)\leftrightarrow (3,4)$ - 
to the expressions in (\ref{su12}). The fact that the reflection matrix for
the $4Q$ dimensional {\sl antisymmetric} representation is diagonal is 
shown first in a study of finite size effects in boundary AdS/CFT 
in \cite{CY} and its implications for the boundary state of the mirror model
are discussed in the Appendix of \cite{BP}.     
 
\section*{Acknowledgments} 

I thank Dr. Z. Bajnok for the discussions and for reading the manuscript. This 
work was supported in part by the Hungarian National Research Fund OTKA under 
K81461.

\end{document}